\documentclass[journal,draftclsnofoot,onecolumn, 12pt]{IEEEtran}

\usepackage{etoolbox}
\newtoggle{1col}
\toggletrue{1col}

\usepackage{etoolbox}

\ifCLASSINFOpdf
   \usepackage[pdftex]{graphicx}
\else
\fi

\usepackage{amssymb}
\usepackage[cmex10]{amsmath}
\usepackage{mathtools}
\usepackage[tight,footnotesize]{subfigure}
\usepackage{booktabs}
\usepackage{color}
\usepackage{paralist}
\usepackage{balance}
\usepackage{bbm}
\usepackage{blindtext}

\usepackage{amsmath}
\usepackage{amsfonts}

\DeclareMathOperator*{\odotoplarge}{\odot}

\begin{document}
\title{SINR Model with Best Server Association for\\ High Availability Studies of Wireless Networks}
\author{David~\"Ohmann, Ahmad Awada, Ingo Viering, Meryem Simsek, Gerhard P. Fettweis}


\IEEEcompsoctitleabstractindextext{
	\begin{abstract}
The signal-to-interference-and-noise ratio (SINR) is of key importance for the analysis and design of wireless networks. For addressing new requirements imposed on wireless communication, in particular high availability, a highly accurate modeling of the SINR is needed. We propose a stochastic model of the SINR distribution where shadow fading is characterized by random variables. Therein, the impact of shadow fading on the user association is incorporated by modification of the distributions involved. The SINR model is capable to describe all parts of the SINR distribution in detail, especially the left tail which is of interest for studies of high availability.
		
	\end{abstract} 
	
}

\maketitle

\IEEEdisplaynotcompsoctitleabstractindextext

\section{Introduction} \label{introduction}
New applications and use cases, introduced in the context of 5\textsuperscript{th} generation (5G) mobile networks, come along with unprecedented and challenging requirements, of which especially high availability is an important cornerstone \cite{NGMN15}. Availability is related to various layers, components, and metrics of wireless communication systems; however, one vital performance indicator, that strongly affects other metrics as well, is the signal-to-interference-and-noise ratio (SINR). 

In contrast to existing studies of the SINR distribution, where typically 5\textsuperscript{th} or 50\textsuperscript{th} percentiles are evaluated, investigations of the left tail of the SINR distribution are required to address the needs of 5G applications, which in turn causes new challenges for system analysis and design. At the left tail of a probability distribution, e.g., at an outage probability of $10^{-7}$ or below, it is difficult to efficiently obtain statistically reliable simulation results. Here, abstract modeling of a system can help address open questions with reasonable complexity. In this work, we present a stochastic SINR model where shadow fading is described by random variables (RVs) shaping the final SINR distribution. The model presented characterizes all parts of the probability distribution in detail, also the leftmost tail.

In this context, user association is an aspect that complicates modeling of the exact SINR distribution. If shadow fading is characterized by RVs, it is non-deterministic which link provides the highest receive power, and hence, the user association is random as well. This is an aspect that is most frequently simplified or neglected in related works. In \cite{Sung10} and \cite{Bulak13}, the authors utilize an approximation for the sum of log-normal RVs and simplify the SINR to a single log-normal RV, but the user association is assumed to be predefined or fixed. Another commonly used simplification is that the serving base station (BS) is chosen based on smallest path loss or shortest distance, see e.g., \cite{Che11}. The latter assumption was also quite common in stochastic geometry; until 2014, when, in \cite{Dhil14}, Dhillon et al. utilized a displacement theorem in order to incorporate shadowing into the user association process. Outside the field of stochastic geometry, there exist only a few approaches that consider shadowing in the user association. In \cite{Muhleis11}, M\"uhleisen et al. present an analysis of the SINR distribution for the Long Term Evolution (LTE) uplink, but it is restricted to only two links. Furthermore, in \cite{Kelif14}, Kelif et al. numerically evaluate the impact of the best server association and show that in certain scenarios it is sufficient to neglect shadowing and connect the users to the closest BS. Another related work can be found in \cite{Minelli13}, where an approximation of the SINR distribution for hexagonal cellular networks is presented. The main idea is to connect the user to a specific BS and then truncate the corresponding SINR distribution below a certain threshold because it is likely that the user connects to another BS. The work in \cite{Oehm15} extends the model from \cite{Minelli13} by shadowing cross-correlation and noise. Although the approaches in \cite{Minelli13} and \cite{Oehm15} are simple and sufficiently accurate for many purposes, e.g., estimating the 50\textsuperscript{th} percentile of the SINR, they are not capable to model the left tail of the SINR distribution with high accuracy.

Our contribution is a new model for analyzing the SINR distribution at specific user locations or over a larger area of an arbitrary but defined cellular deployment. Important features of typical system evaluations such as shadowing cross-correlation and antenna sectorization are considered. We incorporate shadowing into the user association by considering different association options and modification of the power distributions of the interfering links. To elaborate, their distributions are truncated above the power value of the serving link since the latter is always stronger than the interferers. Then, the distributions are summarized to a single SINR distribution by using logarithmic convolution \cite{Punt96}. Most importantly, there is no approximation involved and hence, the model is suited to investigate the left tail of the SINR distribution which can, for instance, be of interest for high availability studies. Finally, we substantiate the accuracy of the model by comparison to Monte Carlo simulations.

\section{SINR Model For Best Server Association} \label{sec_sinr_model}
We describe the SINR distribution in the downlink of a wireless network consisting of $L$ BSs.
A user equipment (UE) at location $m$ receives a signal from BS $j$ with power $P_{m,j}$ (in dBm) given by 
\begin{equation}
P_{m,j} = X_{m,j} + 10 \cdot \log_{10}\left(p_{\text{t},j} \cdot g_{\text{\tiny BS},m,j} \cdot g_{\text{\tiny UE},m,j} \cdot \alpha \cdot d_{m,j}^{-\beta} \right), 
\end{equation}
where $p_{\text{t},j}$ is the transmit power (in mW) of BS $j$, $g_{\text{\tiny BS},m,j}$ and $g_{\text{\tiny UE},m,j}$ are the linear antenna gains of BS $j$ and the UE, respectively, $\alpha$ is the path loss constant, $d_{m,j}$ is the distance between BS $j$ and the UE, and $\beta$ is the path loss exponent. Furthermore, $X_{m,j}$ is a zero-mean Gaussian RV with standard deviation $\sigma_{\text{dB}}$ characterizing random shadowing. Hence, $P_{m,j}$ in turn is a Gaussian RV with the same standard deviation $\sigma_\text{dB}$ and mean 
$\mu_{P_{m,j}} = 10 \cdot \text{log}_{10} \left(p_{\text{t},j} \cdot g_{\text{\tiny BS},m,j} \cdot g_{\text{\tiny UE},m,j} \cdot \alpha  \cdot d_{m,j}^{-\beta} \right)$. In the remainder, probability density functions (PDFs) and cumulative distribution functions (CDFs) are denoted by $f_{(\cdot)}$ and $F_{(\cdot)}$, respectively. Due to space reasons, we focus on the impact of shadow fading and neglect small scale fading. However, if needed, the latter can be added to the final SINR distribution as it is done in \cite{Che11}. Furthermore, we assume a frequency reuse scheme of one and full buffers at the BSs leading to full interference conditions. In the case that the UE is connected to BS $i$, the full interference assumption leads to the following SINR expression
\begin{equation} \label{eq_SINR}
\gamma_{m,i} = \frac{10^{P_{m,i}/10}}{10^{P_{N}/10} + \sum_{j \neq i}10^{P_{m,j}/10}},
\end{equation}
where $P_N$ is the thermal noise power (in dBm) at the receiver. 

\iftoggle{1col}{
	\begin{figure*}[!tb]
		\centering
		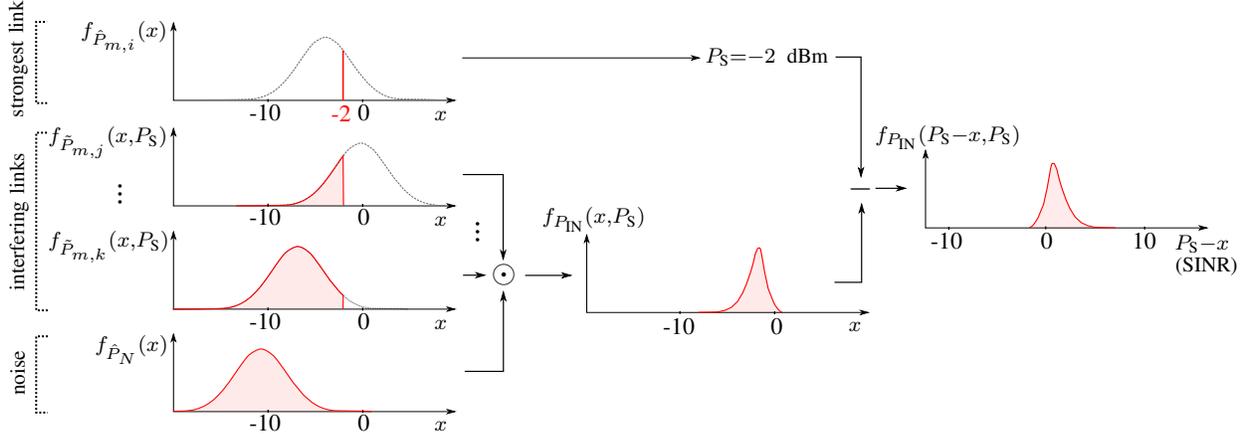
		\caption{Example of the essential procedure of the SINR model. Distributions of the interfering links are truncated above the power value of the serving link. Powers occuring in the denominator of the SINR are then combined by logarithmic convolution.}
		\label{fig_explanation_sinr_model}
	\end{figure*}
}{

}

\subsection{Removal of Shadowing Cross-Correlation}
Since shadowing depends on obstructions, such as buildings and other obstacles, which can be identical for different links, the shadowing processes of individual links are to some extent correlated. In order to consider shadowing cross-correlation in the SINR model, we split the shadowing process into two Gaussian components, according to \cite{Bulak13}, i.e., 
\begin{equation} \label{eq_shad}
X_{m,j} = \sqrt{\rho}\cdot\xi_m + \sqrt{1-\rho}\cdot\eta_{m,j},
\end{equation}
where $\rho$ is the correlation coefficient, $\xi_m$ and $\eta_{m,j}$ are i.i.d. Gaussians, $\xi_m$ describes the impact of the environment close to the UE, and $\eta_{m,j}$ characterizes the link-dependent shadowing with respect to BS $j$. $\xi_m$ is the same for all links while $\eta_{m,j}$ is independent of $\eta_{m,k}, \forall j\neq k$.
Substituting \eqref{eq_shad} into \eqref{eq_SINR} and dividing by the common factor $10^{\left(\sqrt{\rho}\cdot \xi_m\right)/10}$ results in
\begin{equation}
\label{eq_sinr_removed_corr}
\gamma_{m,i}=\frac{10^{\left(\sqrt{1-\rho}\cdot\eta_{m,i} + \mu_{P_{m,i}}\right)/10}}{10^{\left(P_N-\sqrt{\rho}\cdot\xi_m\right)/10} + \sum_{j \neq i} 10^{\left( \sqrt{1-\rho}\cdot\eta_{m,j} + \mu_{P_{m,j}}\right)/10}}.
\end{equation}
Now, all logarithmic power components, hereinafter denoted by $\hat{P}_{m,j}$, have still the same mean values but a reduced standard deviation $\sigma_{\hat{P}_{m,j}}=\sqrt{1-\rho} \cdot  \sigma_\text{dB}$. In addition, the formerly constant noise power has turned into a Gaussian RV $\hat{P}_N$ with mean $\mu_{\hat{P}_N}=P_N$ and standard deviation $\sigma_{\hat{P}_N}=\sqrt{\rho}\cdot \sigma_\text{dB}$. Most importantly, all RVs are independent again which simplifies the following steps.

\subsection{User Association Based on Distance or Path Loss}
In case of a user association based on shortest distance or smallest path loss, the user association is a deterministic process since no random component is involved. Hence, a certain link is chosen as the desired link and the SINR statistics are computed according to \eqref{eq_sinr_removed_corr}, see \cite{Oehm15}. Random shadowing is considered but without having any effect on the user association. 

\subsection{User Association Based on Receive Powers}
However, in real networks, the user association is carried out based on receive powers which are inseparable compounds of path loss and shadowing. So, it is not realistic to consider only geometric properties. Therefore, we model a user association which also depends on random shadowing. 
The SINR for a certain association option is denoted in dB by $\tilde{\Gamma}_{m,i}$
with the distribution $h_{\tilde{\Gamma}_{m,i}}$, subject to the condition that BS $i$ provides the strongest link. Please note that $h_{\tilde{\Gamma}_{m,i}}$, which is explained subsequently, contains a weighting and hence is not a PDF. Adding all distributions that occur for the different association options leads to the final PDF $f_{\tilde{\gamma}_m}$ describing the SINR $\tilde{\gamma}_m$ with best server association, i.e.,
\begin{equation}
\label{eq_weighted_sum_SINR}
f_{\tilde{\gamma}_m} (x) = \sum_{i=1}^{L} h_{\tilde{\Gamma}_{m,i}}\left(10\cdot \log_{10} \left(x\right)\right).
\end{equation}

The SINR distribution $h_{\tilde{\Gamma}_{m,i}}$ is derived based on the constraint that link $i$ is chosen as the desired link which means that no other link is stronger than link $i$. This constraint also influences the receive powers in the denominator of the SINR. For each possible receive power of link $i$, we compute the power sum of the interfering links where the power distributions of the interfering links are modified. For instance, in case link $i$ experiences a receive power $\hat{P}_{m,i}=P_\text{S}$, the power distribution of an interfering link $j$ is truncated accordingly, i.e.,
\begin{equation}
\label{eq_interf_power}
f_{\tilde{P}_{m,j,\text{omni}}} (x,P_\text{S}) = 
\left\{
\begin{array}{ll}
\frac{f_{\hat{P}_{m,j}}(x)}{\int_{-\infty}^{P_\text{S}} f_{\hat{P}_{m,j}}(y) \text{d}y}  & \text{for} \,\, x \leq P_\text{S},
\\0  & \text{for} \,\, x > P_\text{S}.
\end{array}
\right.
\end{equation}

In case of omnidirectional antennas, the result of \eqref{eq_interf_power} can be directly used in the following steps, i.e., 
\begin{equation} \label{eq_diff_ant_types}
f_{\tilde{P}_{m,j}}(x,P_\text{S}) = f_{\tilde{P}_{m,j,\text{omni}}}(x,P_\text{S}). 
\end{equation}
For sectorized antennas, we use a different assignment of $f_{\tilde{P}_{m,j}}(x,P_\text{S})$ which is explained later in Sec.~\ref{sec_sect_ant}.

Next, the resulting PDFs of the interfering powers and the unaffected PDF of the noise power are combined to a joint power distribution. Unfortunately, no closed form solution exists for the sum of logarithmic powers. Even for the case of standard log-normal distributions, no simple closed form solution is known, see \cite{Sung10}, \cite{Bulak13}. However, according to \cite{Punt96}, the PDF of the sum of receive powers with arbitrary PDFs can be derived by the so-called \textit{logarithmic convolution} by
\begin{align}
\label{eq_log_conv}
f_R (r) =&  \{f_X \odot f_Y\}(r) = \int_{-\infty}^{r} f_X(z) \cdot f_Y\left(D(r,z)\right) \text{d}z&  \nonumber\\
& \quad\quad\quad\quad\quad +  \int_{-\infty}^{r} f_X\left(D(r,z)\right) \cdot f_Y(z) \, \text{d}z,  &
\end{align}
with $D(r,z) = 10\cdot \log_{10} \left(10^{r/10}-10^{z/10}\right)$. $f_R$, $f_X$, and $f_Y$ denote PDFs of logarithmic powers,  and $\odot$ is introduced as an operator for logarithmic convolution. In case of more than two terms, the convolution is applied recursively. We compute the joint distribution of the sum of all powers in the denominator of the SINR expression by 
\begin{equation}
\label{eq_inter_pow}
f_{P_\text{IN}}(x,P_\text{S}) = \{(\odotoplarge_{j\neq i} f_{\tilde{P}_{m,j}}) \odot f_{\hat{P}_{N}}\}(x,P_\text{S}).
\end{equation}
The numerator of the SINR is a scalar and hence the SINR distribution can be derived by simple subtraction. This is done for all possible receive powers of link $i$ which leads to
\begin{equation}
\label{eq_BSP_one_case}
h_{\tilde{\Gamma}_{m,i}}(x) =  \int_{-\infty}^{+\infty} \omega_i(P_\text{S}) \cdot  f_{P_\text{IN}}(P_\text{S}-x,P_\text{S}) \, \text{d}P_\text{S},
\end{equation}
where $\omega_i(P_\text{S})$ describes the probability that BS $i$ provides the strongest link, given by 
$\omega_i(P_\text{S}) = f_{\hat{P}_{m,i}}(P_\text{S}) \cdot \prod_{j \neq i} F_{\hat{P}_{m,j}}(P_\text{S})$. The crucial part of the procedure is illustrated in Fig.~\ref{fig_explanation_sinr_model}. Finally, the aforementioned steps are performed for each link, and by using \eqref{eq_weighted_sum_SINR}, the final SINR distribution is derived.

\subsection{Extension to Sectorized Sites} \label{sec_sect_ant}
So far, the model is tailored to a single BS per site, but sectorization can be added as follows. The sectors of site $j$ are numbered by $s \in \mathcal{S}_j = \{1,...,S_j\}$ with $S_j$ being the number of sectors of site $j$. The antenna gain $g_{\text{\tiny BS},m,j}$ is replaced by the antenna pattern gains $g_{\text{\tiny BS},m,j,s}$. Furthermore, we consider the strongest sector of each site $s^{*}$ and its corresponding antenna pattern gain $g_{\text{\tiny BS},m,j,s^{*}}$ in all expressions in Sec.~\ref{sec_sinr_model}. Please note that sectors of the same site experience the same shadowing ($\rho=1$) and hence cannot be regarded as standard BSs. We consider the additional sectors as follows: (i) If all sectors of site $j$ are interfering, we replace \eqref{eq_diff_ant_types} by $f_{\tilde{P}_{m,j}}(x,P_\text{S}) = f_{\tilde{P}_{m,j,\text{omni}}}(x-G_\text{sec},P_\text{S})$, where $G_\text{sec}$ describes the additional interfering power of the weaker sectors given by
$G_\text{sec} = 10\cdot \log_{10} \left(1+\sum_{s\in \mathcal{S}_j \setminus \{s^{*}\}} g_{\text{\tiny BS},m,j,s} / g_{\text{\tiny BS},m,j,s^{*}}\right)$.
(ii) Furthermore, there are interfering sectors at the site of the serving sector. Their sum power is given by $P_\text{S} + 10\cdot \log_{10}  \left( 
\sum_{s\in \mathcal{S}_i \setminus \{s^{*}\}}
  g_{\text{\tiny BS},m,i,s}/g_{\text{\tiny BS},m,i,s^{*}}\right)$ which is added to $P_\text{IN}$ after computing \eqref{eq_inter_pow}.

\subsection{Aspects of Implementation}
Unfortunately, no closed form solution exists for the logarithmic convolution. Hence, we propose to solve \eqref{eq_inter_pow} by numerical integration. Other dependent expressions such as \eqref{eq_BSP_one_case} are evaluated numerically as well. For doing this, the receive powers are discretized with a certain granularity and within an appropriate value range\footnote{In the numerical evaluation presented subsequently, the receive powers are discretized in 0.1~dB steps between $P_\text{max}-80$~dB and $P_\text{max}+40$~dB, where $P_\text{max}$ is the maximum mean receive power occurring at a certain user location.}. Then, the procedure indicated in Fig.~\ref{fig_explanation_sinr_model} is executed for each of those receive powers. The accuracy of a single numerical integration depends on the integration method, the function to be integrated, and the granularity of the discretization. We employ a trapezoidal integration which exhibits a second-order error bound. The errors of the recursive integrations add up, and hence, it is important to choose an appropriate trade-off between the integration method, the number of integrations, and the granularity. This trade-off also affects the computational complexity of the computation which grows with the number of links and the granularity of the discretization. Please note that the complexity of the SINR model is basically independent of the outage probability of interest. In contrast, the complexity of the Monte Carlo simulation performed in Sec.~\ref{sec_numer_eval} is linearly dependent upon the number of simulation runs which in turn can be adjusted according to the outage probability of interest. For the implementation and settings used in this work, the model shows similar execution time to a simulation with $10^{7.5}$ simulation runs. Such a sample space is sufficient to study, for instance, an outage probability of $10^{-5}$ with a 99\% confidence interval of width $ \pm 10^{-6}$. However, if considerably more simulation runs are required because much lower outage probabilities are of interest, using the SINR model is more computationally efficient than performing extensive simulations. These observations clearly underline the necessity of an SINR model for studying extremely low outage probabilities.

\iftoggle{1col}{
	\begin{figure}[!b]
		\centering
		\includegraphics[width=0.51\linewidth]{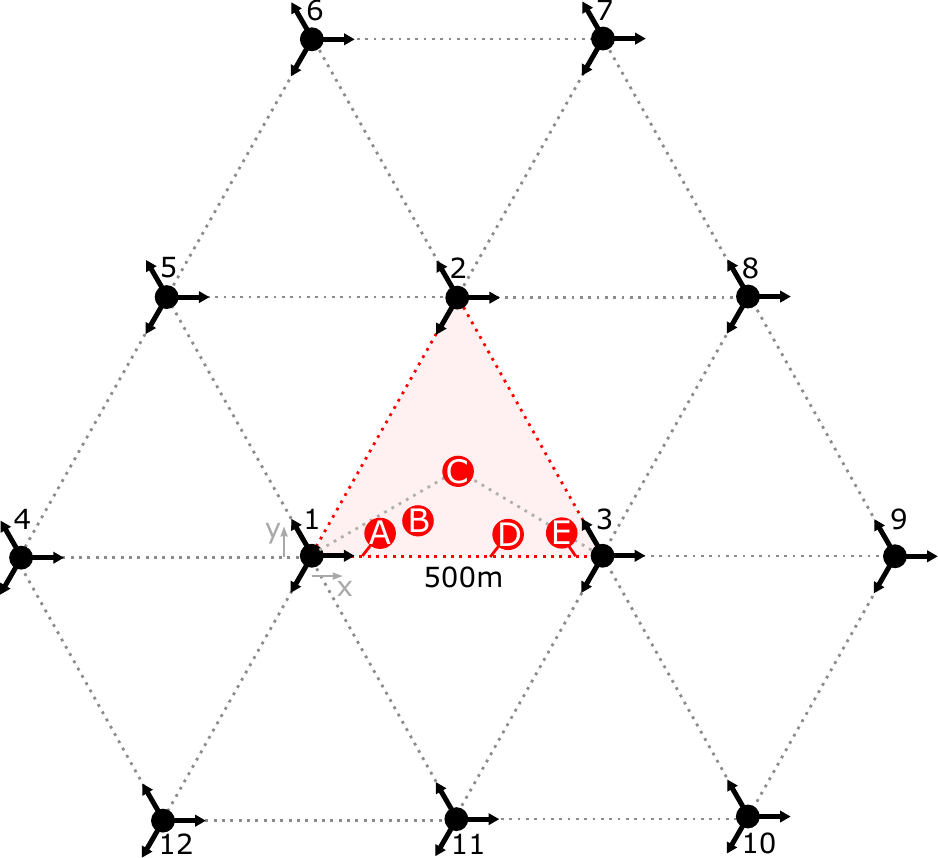}
		\caption{Evaluation scenario consisting of 12 sites. The inner triangle evaluated is marked in red. The exemplarily studied user locations are also indicated.}
		\label{fig_evaluation_scen}
	\end{figure}
}{

\begin{figure}[!b]
	\centering
	\includegraphics[width=0.650\linewidth]{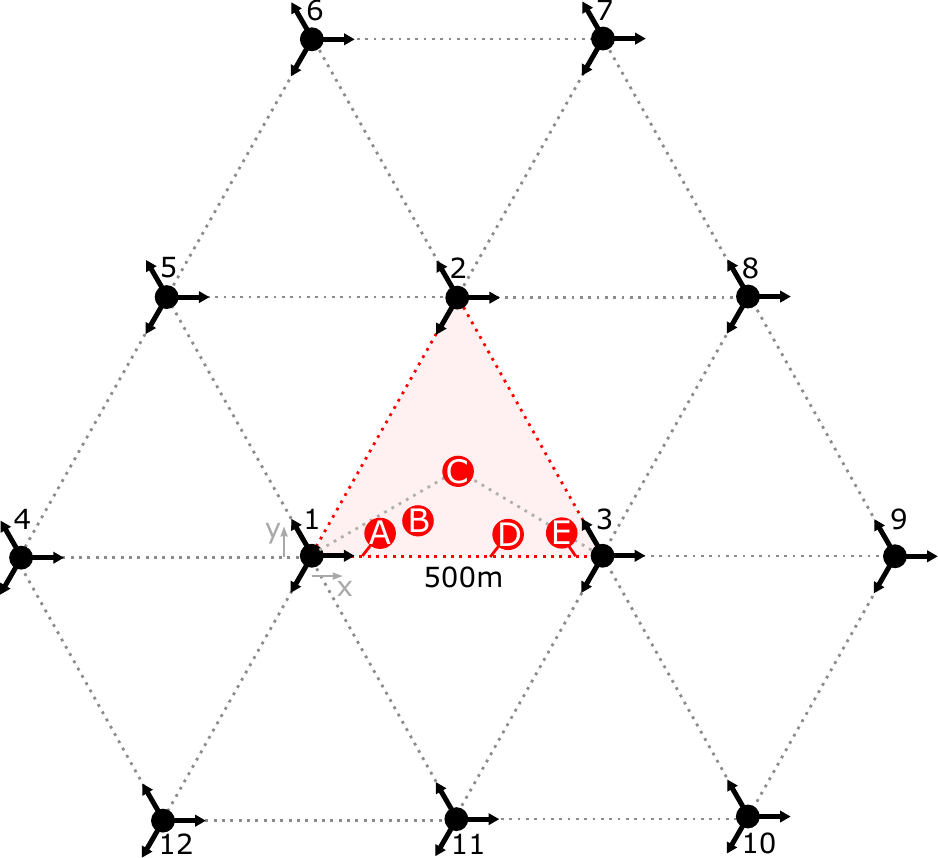}
	\caption{Evaluation scenario consisting of 12 sites. The inner triangle evaluated is marked in red. The exemplarily studied user locations are also indicated.}
	\label{fig_evaluation_scen}
\end{figure}
}

\iftoggle{1col}{
	\begin{figure}[!tb]
		\centering 
		\subfigure[CDF of the SINR in linear scaling.]{\label{fig_res_perf1}\includegraphics[width=0.62\textwidth]{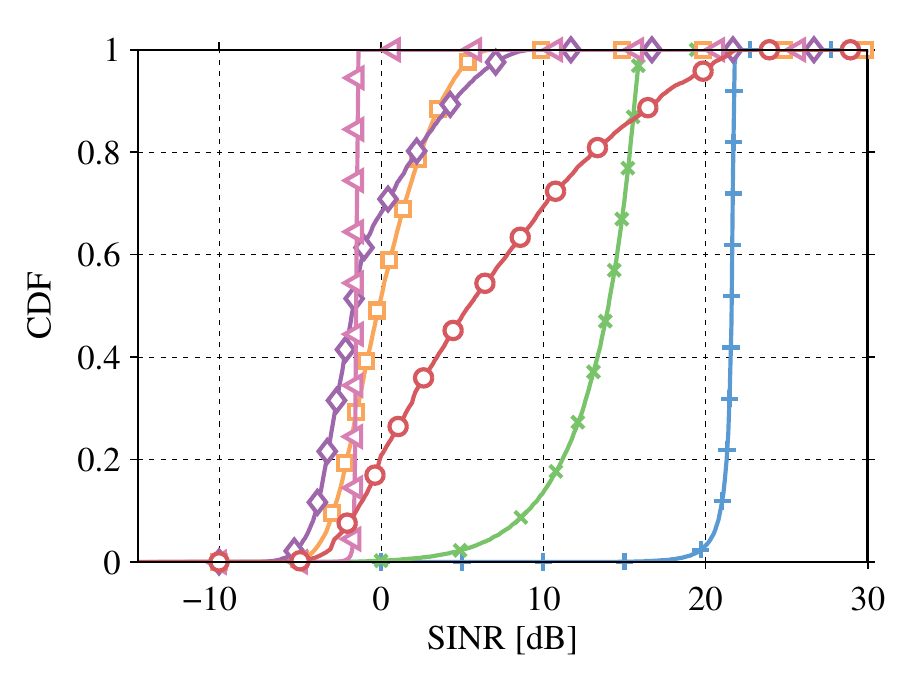}}
		\subfigure[CDF of the SINR in logarithmic scaling.]{\label{fig_res_perf2}\includegraphics[width=0.62\textwidth]{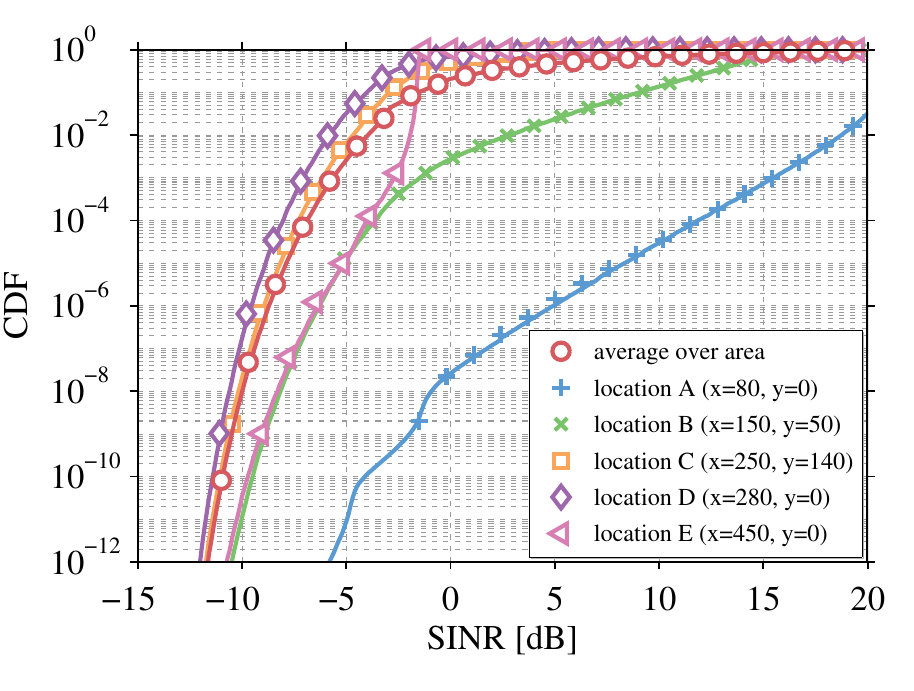}}
		\caption{Numerical evaluation of the SINR model: Model (solid lines) and Monte Carlo simulations (markers). The legend of Fig.~\ref{fig_res_perf2} also applies to Fig.~\ref{fig_res_perf1}.}
		\label{fig_res_perf_large}
	\end{figure}
}{

}

\section{Numerical Evaluation} \label{sec_numer_eval}
In this section, we evaluate the accuracy of the SINR model by comparing it to Monte Carlo simulations in a realistic LTE environment. As depicted in Fig.~\ref{fig_evaluation_scen}, twelve three-fold sectorized sites are arranged in a hexagonal grid. 

The simulation parameters, e.g., channel characteristics and 3D antenna patterns, are set according to typical LTE system evaluation settings \cite{3GPP_36814}. Distance-dependent path loss is given by $128.1+ 37.6\cdot \log_{10}(d)$, where $d$ is given in km. A carrier frequency of $2$~GHz, a bandwidth of $20$~MHz, and a transmit power per BS of $49$~dBm are used. Furthermore, the shadowing standard deviation is set to $\sigma_\text{dB}=8$~dB. The shadowing correlation between sites is $\rho=0.5$ and the correlation between sectors of the same site is $\rho=1$.

The inner triangle marked in light red is evaluated in a 10~m grid. Assuming a homogeneous user distribution $\delta_m$ over the triangle area $\mathcal{A}$, an overall SINR distribution can be computed by $f_{\tilde{\gamma}} (x) = \int_{\mathcal{A}} \delta_m \cdot f_{\tilde{\gamma}_m} (x)\,\text{d}m$. This overall SINR distribution characterizes the performance experienced by a randomly selected UE in such a system setup without knowledge about the actual development and related shadowing conditions.

The SINR distributions for the overall area and also for some exemplary user locations, which are marked in Fig.~\ref{fig_evaluation_scen}, are depicted in linear and logarithmic scaling in Fig.~\ref{fig_res_perf_large}. The solid lines describe the results of the SINR model while the markers indicate the results of Monte Carlo simulations each with $10^9$ shadowing realizations. 
The results of the Monte Carlo simulations corroborate the modeling results up to an outage probability of approximately $10^{-9}$. Since the courses of the curves do not exhibit any abrupt changes, we believe that the model is also accurate for lower outage probabilities. Moreover, in this specific system setup, the SINR distributions are quite different for varying user locations. For instance, the 50\textsuperscript{th} percentile of the SINR ranges from $-2$~dB (location D) to $22$~dB (location A). Furthermore, Fig.~\ref{fig_res_perf2} indicates how the SINR model can help analyze the availability of a network. In the scenario at hand, a minimum SINR of $-8$~dB is achieved with a probability of approximately $1-10^{-5}$ which might be reliable enough for certain applications, for others not, cf. \cite{NGMN15}. Please note that other causes of outage, e.g., mobility and small scale fading, may increase the outage probability.

\section{Conclusion}
As a toolset for fine-grained studies of SINR distributions, we presented an SINR model capturing shadowing, the impact of shadowing on the user association, shadowing cross-correlation, and antenna sectorization. A numerical evaluation substantiated that all parts of the SINR distribution are described with high accuracy. 

On this basis, the model is well suited for upcoming high availability studies of wireless networks. The model can be applied to heterogeneous networks by using specific transmit powers, path loss models, and antenna settings for individual base stations, e.g., low-power small cells with omnidirectional antennas and high-power macro base stations with sectorized antennas. Furthermore, the model can be extended by beamforming techniques in order to evaluate higher carrier frequencies, e.g., millimeter wave,  enabling detailed comparison of different frequency layers discussed for 5G networks.

\end{document}